\documentstyle[twocolumn,aps,prl,epsf,floats]{revtex}
\begin{document} 
\title {On the vertex corrections in antiferromagnetic 
spin fluctuation theories}
\author{ Andrey V. Chubukov$^{1,2}$, Philippe Monthoux$^{3}$, 
Dirk K. Morr$^{1}$ }
\address{
$^{1}$ Department of Physics, University of Wisconsin, Madison, WI 53706 \\
$^{2}$ P.L. Kapitza Institute for Physical Problems, Moscow, Russia \\
$^{3}$ Department of Physics and National High Magnetic Field Laboratory,
Florida State University, Tallahassee, FL 32306}
\date{today}
\maketitle

\begin{abstract}
We argue that recent calculations by Amin and Stamp
 (PRL {\bf 77}, 301 (1996)) 
overestimate the strength
of the vertex corrections in the spin-fermion model for cuprates.
 We clarify the physical origin of the apparent 
discrepancy between their results and earlier calculations. 
We also comment on the relative sign of the vertex correction.
\end{abstract}


In a recent publication~\cite{paper}  Amin and Stamp 
computed the leading order correction to the spin-fermion vertex in the nearly
antiferromagnetic Fermi-liquid model. They considered a 
region near optimal doping in which the Fermi surface is large, and   
precursors of the spin-density-wave state have not been formed yet. 
They argued that the leading vertex correction is rather large ($\Delta g/g
\sim 1$) for realistic values of the parameters of the 
electron dispersion and the spin fluctuation spectrum. 
This result is in clear 
disagreement with previous calculations \cite{andr,phil,AIM} which 
reported a much smaller 
amplitude of the correction. Whether vertex corrections are large or not
 is relevant to the validity of the
Eliashberg-type calculations of
the superconducting transition temperature in cuprates~\cite{pines}.

In this letter, we clarify the origin of the discrepancy between Ref. [1] and
earlier results and show that  Ref. [1] strongly overestimates
the  amplitude of the vertex correction  near optimal doping.

The spin-fermion model describes fermions coupled to spin fluctuations by 
\begin{equation}
{\cal H}_{s-f} = g 
\sum c^{\dagger}_{k,\alpha} {\vec \sigma}_{\alpha,\beta}
c_{k+q,\beta} {\vec S}_{-q}.
\label{1}
\end{equation}
Here $g$ is the coupling constant, and $\sigma_i$ are the Pauli matrices. 
It is assumed that the Fermi-liquid description is valid, i.e., near the Fermi
surface, the electronic Green's function behaves as
$G(k, \omega_m) = Z/(i \omega_m - {\bar\epsilon}_k)$, where ${\bar\epsilon}_k =
\epsilon_k - \mu$, and $Z \leq 1$ is a positive
constant. The dispersion near the Fermi surface is given by
$\epsilon_k = -2t (\cos k_x + \cos k_y) -4t^{\prime} \cos k_x \cos k_y$. 
Spin fluctuations are described by a dynamical spin
susceptibility which is assumed to be strongly peaked near the
antiferromagnetic momentum $Q=(\pi,\pi)$, and to behave at low energies as
$\chi (q,\omega_m) = \chi_Q/(1 + \xi^2 {\tilde q}^2 + i
|\omega_m|/\omega_{sf})$. Here ${\tilde q} = q-Q,~\xi$ is the magnetic
correlation length, and $\omega_{sf} \propto \xi^{-2}$ 
is a typical spin fluctuation frequency
which is much smaller than any other 
energy scales in the problem due to the proximity to antiferromagnetism.
Amin and Stamp computed the vertex corrections for the points on the Fermi
surface which are connected by Q (the so-called ``hot spots''), and also for
$k$ along the Brillouin zone diagonal. They argued that the relative correction is
larger for the hot spots - a result we do not dispute. 
For the vertex correction at the hot spot they obtained 
{\it assuming that fermions are free particles}
$\Delta g/g =  - (g^2 \chi_Q \omega_{sf}/4 \pi^3 \mu^2) I (k_{h})$ where 
$I (k_{h})$ is a dimensionless quantity. It turns out that $I(k_{h})$ is rather 
large which implies a large amplitude of the correction. 
Amin and Stamp claim that the origin of the discrepancy between their and earlier 
results~\cite{andr,phil,AIM} lies in their more accurate numerical evaluation of 
$I (k_h)$. We disagree with their explanation 
and will argue that the reason for their
discrepancy with earlier results in fact has a physical 
rather than a numerical origin.
 
\begin{figure}[t]
\begin{center}
\leavevmode
\epsffile{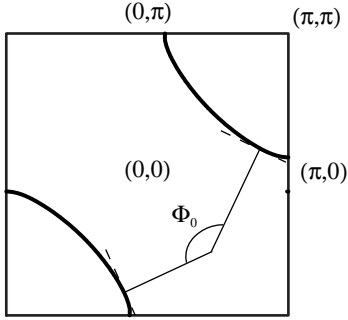}
\end{center} 
\caption {The graphical representation of the angle $\phi_0$ 
between the normals to the Fermi surface 
at hot spots (dashed lines). 
For clarification we omitted the parts of the Fermi surface in 
the second and fourth quadrant.}
\label{phi0}
\end{figure}

We independently computed the vertex corrections analytically and 
numerically along the same lines as described above. The analytical 
expression for $\Delta g/g$ 
 is given by
\begin{eqnarray}
\frac{\Delta g}{g} &=& - \frac{g^2 Z^2\chi_Q \omega_{sf}}{ \pi^3 v^2}
\times \nonumber\\
&&\left[~Re \int_0^{\pi}~d\phi~
\frac{\log[\sin (\phi/2)]}{\cos \phi + \cos \phi_0} \log\frac{\sin
(\phi/2)}{\delta^2} + O(\delta^2)\right]
\label{beta}
\end{eqnarray}
where $\delta = c^2_{sw}/(2\gamma~v~\xi) \equiv
 \xi \omega_{sf}/v \ll 1$, and $v$ and $\phi_0$ 
define the fermionic dispersion around a hot spot:
$\epsilon_k = v(k-k_{h}) \cos (\phi)$ and $\epsilon_{k+Q} = v(k-k_{h}) 
\cos (\phi + \phi_0)$. The angle $\phi_0$ is the angle between the normals
to the Fermi surface at the hot spots (see the Fig.\ref{phi0}).

We emphasize that Eq. (\ref{beta}) contains a term which logarithmically
depends on $\delta$ {\it and} a term independent of $\delta$. 
Previous analytical calculations of the vertex correction~\cite{andr}
 restriced
with the $|\log \delta|$ term only. We will see that for parameters relevant to
cuprates both terms nearly equally contribute to the vertex renormalization.

We first compare our analytical (Eq.(\ref{beta})) 
and numerical~\cite{phil} results
with the ones obtained by Amin and Stamp for the same set of parameters,
namely $t=0.25 eV,~t^{\prime} =-0.45 t,~|\mu| =1.46 t,~\omega_{sf} =
14 meV,~\xi =2.5a,~g=0.64 eV$ and  $Z=1$. They found
$\Delta g/g \approx -0.7$  while Eq.(\ref{beta}) yields 
$\Delta g/g \approx 0.55$, and numerical computations~\cite{phil}
yield $\Delta g /g \approx 0.4$ (for this set of parameters $\Phi_0=1.78$
 and $\delta=0.27$). 
We see that apart from the sign difference which we discuss
later, our results and the ones by Amin and
Stamp are in close agreement which implies that a more accurate evaluation of 
$I(k_h)$ {\it cannot} be the main reason for the discrepancy mentioned above.  
The actual reason for the discrepancy with earlier work
lies in the fact that Amin and Stamp considered $\omega_{sf}$ as an input
parameter, whose value can be inferred from NMR experiments, and simultaneously
treated fermions as free particles with $Z=1$ (!). We will show that these
two assumptions are likely to be incompatible.

The point is that near optimal doping,  the key source for the 
spin damping is the interaction
with fermions (the damping due to this interaction is much stronger than the one due to a
direct spin-spin exchange). 
Since the damping term in the spin susceptibility is related to the imaginary part of 
the particle-hole bubble at momentum transfer $Q$ and low frequencies,  
the fermions which contribute to $\omega_{sf}$ are 
located in the vicinity of the hot spots. One can then compute the damping within the
spin-fermion model of Eq.(\ref{1}) 
which, we remind, is valid only near the Fermi surface where fermionic
excitations are coherent. The result is
\begin{equation}
\omega_{sf} = \frac{\pi}{4}~|\sin \phi_0|~ \frac{v^2}{g^2 Z^2 \chi_Q}.
\label{osf}
\end{equation}
If one substitutes the same values for the parameters as above, one obtains 
$\omega_{sf} \approx 1.06 meV$ which is more than ten times less 
than $\omega_{sf} =14 meV$  used in the above calculations of $\Delta g/g$. 
A way to restore the experimentally inferred value of $\omega_{sf}$ is to assume
that the fermions are not free particles, 
i.e., $Z<1$. Specifically, one needs $Z = 0.28$ to obtain the 
correct value of $\omega_{sf}$. Substituting $Z=0.28$ into the formula
for $\Delta g/g$ (Eq.(\ref{beta})), we immediately 
obtain $\Delta g/g \approx +0.04$.
Note in passing that a small value for $Z$ is consistent with  photoemission
data ~\cite{photo} which show that even at optimal doping, 
the quasiparticle peak is rather broad, and there is a substantial spectral weight in the
incoherent part of the spectrum.

Alternatively, one can consider fermions with arbitrary $Z$, 
compute $\omega_{sf}$  using Eq.(\ref{osf}),
and substitute the result into Eq.(\ref{beta}). 
Doing this, we obtain
neglecting terms of $O(\delta^2)$ 
\begin{equation}
\frac{\Delta g}{g} = - \frac{|\sin \phi_0|}{4\pi^2}
Re \int_0^{\pi}~d\phi~
\frac{\log[\sin (\phi/2)]}{\cos \phi + \cos \phi_0} \log\frac{\sin
(\phi/2)}{\delta^2}
\label{beta2}
\end{equation}
This is the way the vertex correction was obtained in \cite{andr}. 
For the set of parameters chosen, we indeed recover the same result.
Similar calculations for $\phi_0 =\pi$ have been 
performed in Ref.\cite{AIM} which also yielded a very small value of 
$\Delta g/g$. 

We now address two technical points.
The first one is the  sign of the vertex correction. 
We have already mentioned that  the sign obtained in 
earlier calculations~\cite{andr,phil,AIM} 
is opposite to the one obtained by Amin and Stamp. 
A simple way to check the sign of the correction
is to consider the limit where both the coupling constant and the 
chemical potential are much larger than the fermionic bandwidth, and $|\mu|
\approx g/2$. In this limit
the electronic structure develops precursors of the spin-density-wave state, as
two of us have recently demonstrated explicitly~\cite{ChuMorr}.
Accordingly,
the renormalized spin-fermion vertex should be much smaller than the bare
one due to a ``near'' Ward identity (our argument here parallels the one 
recently displayed by Schrieffer~\cite{Schrieff}). 
Meanwhile, a simple examination of Eq.(6) in Ref.\cite{paper}
shows that at large $\mu$, the vertex correction  is
{\it positive}, i.e., $\Delta g/g \rightarrow +1$ rather than tending to $-1$
which is necessary to obtain the  physically motivated strong reduction of the
vertex.
 
The second point concerns the computations near
 the edge of the antiferromagnetic
instability, when the correlation length is very large.
 It follows from Eq.(\ref{beta}) that  
as $\delta \propto \omega_{sf} \xi \propto \xi^{_1}$ tends to zero, 
$\Delta g$ diverges logarithmically. 
In this situation, higher-order
corrections are indeed relevant. It has been shown in~\cite{andr,AIM}
that the logarithms  sum up to a power law, and the full vertex takes the form
\begin{equation}
g_{tot} = g \left(\frac{\xi}{a}\right)^{\beta}
\end{equation}
where 
\begin{equation}
\beta = \frac{|\sin \phi_0|}{2\pi^2}
Re \int_0^{\pi}~d\phi~
\frac{\log[\sin (\phi/2)]}{\cos \phi + \cos \phi_0} 
\end{equation}
As we already discussed, however, $\beta$ is negative and numerically quite
small.
We thank  D. Pines and P.C.E. Stamp for useful conversations. 
The work by A. Ch. and D. M. has been supported by the NSF DMR 9629839. A. Ch.
is an A.P. Sloan Fellow. P.M. acknowledges support from the National High 
Magnetic Field Laboratory and the State of Florida.

\end{document}